\documentclass[11pt,a4paper]{article}
\usepackage{amsmath}
\usepackage{amssymb}

\begin{document}

\title{ \bf A new approach in theoretical physics based on the Einstein
covariance principle}
\author{\textbf{V.M. Mekhitarian and V.E. Mkrtchian} \\
Institute for Physical Research, Armenian Academy of Sciences, \\
Ashtarak-2, 378410, Republic of Armenia.}
\maketitle

\begin{abstract}
In this article we show that Einstein covariance principle provides a wide
opportunity in the solutions of different problems of theoretical physics.
Here we apply covariance principle in some problems of classical
electrodynamics and kinetics. Extension of this approach in the other fields
is obvious.
\end{abstract}

\section{INTRODUCTION}

The requirement of general covariance of the equations describing different
processes in the nature is one of the corner-stones of Einstein general
relativity$^{1}$and has enormous significance in modern theoretical physics.
With the equivalence principle it brings gravitation to the metric
properties of the space-time and shows connection of the geometry of space
-time continuum with the material processes.

In this article we are going to open another powerful aspect of the Einstein
covariance principle. Namely, we are going to show that covariance principle
provides new methods for the solutions of different problems of theoretical
physics.

Here we will apply principle of covariance to some problems of classical
electrodynamics and kinetics. Extension of this approach to the other fields
of physics is obvious.

The prehistory of this work is very short. More than twenty years ago one of
the authors found out importance of Euler transformation in the classical
electrodynamics$^{2}$. Then, with use of the covariance principle, the
boundary-value problem of the electrodynamics has been solved for an
expanding-contracting sphere$^{3}$.

The paper is organized as follows. In Section 2 the Euler transformation is
introduced, notations are given and preliminary explanations of the
mathematical procedure in the article. In Section 3, with the help of
covariance principle, we construct infinitely many solutions of the
continuity equation from the given one. In Section 4 it is shown that the
Euler transformation provides an opportunity to express charge and current
densities of the ensemble of point particles in any arbitrary field as a
linear combination of the same quantities in the absence of the external
field. Application of the covariance principle in the Boltzman kinetic
equation is given in Section 5. It is shown here that the Euler
transformation enables one to find out a solution of the nonrelativistic
Boltzman equation in any external field if a solution of the same equation
in the absence of the external field is known. Sections 6,7 are concerned
with the problems of Maxwell's phenomenological electrodynamics. As an
illustration of the developed technique we give a solution of the problem of
propagation of electromagnetic waves in an inhomogeneous anisotropic medium.
In Section 8 we consider a self- consistent problem of plasma
electrodynamics. This Section shows the way of construction an exact
solution of the interaction of electron plasma with the external
electromagnetic field on the slab of positive ions.

\section{BACKGROUND}

\textbf{a. }In calculations we will use notations of the well known,
Landau-Lifshitz book$^{4}$.

\textbf{b. }Throughout the article we will follow the same procedure,
i.e.,we will consider an equation having some solution in an inertial
reference frame $K^{\prime }$ with Cartesian coordinates $X^{\prime i}\left(
ct^{\prime },\mathbf{r}^{\prime }\right) $ and metric tensor%
\begin{equation}
g^{^{\prime }ij}=g_{ij}^{\prime }=\left( 
\begin{array}{c}
1000 \\ 
0-100 \\ 
00-10 \\ 
000-1%
\end{array}%
\right) .  \label{1}
\end{equation}

Then we perform an arbitrary transformation 
\begin{equation}
X^{\prime i}=W^{i}\left( X\right) ,\text{ \ }i=0,1,2,3.  \label{2}
\end{equation}

to a noninertial reference frame $K$ with the covariant form of given
equation. Just comparing the forms of the equation in $K^{\prime }$ and $K$
we find a new set of solutions of the given equation in the inertial frame $%
K^{\prime }.$

\textbf{c. }In the noninertial frame $K$ covariant components of metric
tensor are%
\begin{equation}
g_{ij}=\Lambda _{i}^{m}\Lambda _{j}^{n}g_{mn}^{\prime },  \label{3}
\end{equation}

where

\begin{equation}
\Lambda _{j}^{i}\left( X\right) =\frac{\partial X^{\prime i}}{\partial X^{j}}%
\equiv \partial _{j}W^{i}\left( X\right)  \label{4}
\end{equation}

is the matrix of transformation ( \ref{2}) .

The contravariant components of metric tensor of $K$ is given by expression

\begin{equation}
g^{ij}=\tilde{\Lambda}_{m}^{i}\tilde{\Lambda}_{n}^{j}g^{\prime mn},  \label{5}
\end{equation}

where $\tilde{\Lambda}$ is the reciprocal matrix of $\Lambda $%
\begin{equation}
\Lambda _{m}^{i}\tilde{\Lambda}_{j}^{m}=\tilde{\Lambda}_{m}^{i}\Lambda
_{j}^{m}=\delta _{j}^{i}.  \label{6}
\end{equation}

\textbf{d. }We use Euler transformation which is very well known in the
theory of elasticity as

\begin{equation}
\mathbf{r}^{\prime }\mathbf{=r-u}\left( \mathbf{r,}t\right)  \label{7}
\end{equation}

where $\mathbf{u}\left( \mathbf{r,}t\right) $ is the displacement field of
medium. Here we will consider four-dimensional form of (7): 
\begin{eqnarray}
X^{\prime i} &=&X^{i}-U^{i}\left( X\right) ,  \label{8} \\
U^{0} &=&0.  \notag
\end{eqnarray}

\textbf{e. }With the help of ( \ref{4})  and ( \ref{8})  we
have the following expression of $\Lambda $, $\tilde{\Lambda}$ for Euler
transformation:%
\begin{equation}
\Lambda _{0}^{0}=1,\text{\ }\Lambda _{\alpha }^{0}=0,\text{ }\Lambda
_{0}^{\alpha }=-\frac{\dot{u}_{\alpha }}{c},\text{ }\Lambda _{\beta
}^{\alpha }=\delta _{\alpha \beta }-\partial _{\beta }u_{\alpha }\equiv
S_{\alpha \beta },  \label{9.a}
\end{equation}

\begin{equation}
\tilde{\Lambda}_{0}^{0}=1,\text{ }\tilde{\Lambda}_{\alpha }^{0}=0,\text{ }%
\tilde{\Lambda}_{0}^{\alpha }=\frac{\dot{u}_{\beta }}{c}S_{\alpha \beta
}^{-1},\text{ }\tilde{\Lambda}_{\beta }^{\alpha }=S_{\alpha \beta }^{-1}. 
\label{9.b}
\end{equation}

Here $u_{\alpha }$ $\left( \alpha =1,2,3\right) $ are components of $\mathbf{%
u}$, $S_{\alpha \beta }^{-1}$ is reciprocal matrix of $S_{\alpha \beta }$,
and 
\begin{equation*}
\frac{\dot{u}_{\alpha }}{c}\equiv \partial _{0}u_{\alpha }.
\end{equation*}

Using ( \ref{3}) , ( \ref{5}) and (\ref{9.a}), (\ref{9.b}) we have
for the covariant and contravariant components of the metric tensor in
reference frame $K$, respectively%
\begin{equation}
g_{00}=1-\frac{\mathbf{\dot{u}}^{2}}{c^{2}},\text{ }g_{0\alpha }=\frac{\dot{u%
}_{\beta }}{c}S_{\beta \alpha },\text{ }g_{\alpha \beta }=-S_{\lambda \alpha
}S_{\lambda \beta },  \label{10.a}
\end{equation}

\begin{equation}
g^{00}=1,\text{ }g^{0\alpha }=\frac{\dot{u}_{\beta }}{c}S_{\alpha \beta
}^{-1},\text{\ }g^{\alpha \beta }=-S_{\alpha \sigma }^{-1}S_{\beta \lambda
}^{-1}(\delta _{\sigma \lambda }-\frac{\dot{u}_{\sigma }\dot{u}_{\lambda }}{%
c^{2}}).  \label{10.b}
\end{equation}

Then, for the determinant of covariant metric tensor we get: 
\begin{equation}
\sqrt{-g}=1-\partial _{\lambda }a_{\lambda },  \label{11.a}
\end{equation}

\begin{equation}
a_{\lambda }=u_{\lambda }+\frac{1}{2}\left[ u_{\nu }\partial _{\nu
}u_{\lambda }-u_{\lambda }\partial _{\nu }u_{\nu }\right] +\frac{1}{3}\sigma
_{\lambda \nu }u_{\nu },  \label{11.b}
\end{equation}

\begin{equation}
\sigma _{\alpha \beta }=\frac{1}{2}e_{\alpha \mu \nu }e_{\beta \rho \lambda
}(\partial _{\mu }u_{\rho })(\partial _{\nu }u_{\lambda }).  \label{11.c}
\end{equation}

Space components of contravariant metric tensor are determined in (\ref{10.b}) via expressions:%
\begin{equation}
\sqrt{-g}S_{\alpha \beta }^{-1}=\delta _{\alpha \beta }-\partial _{\nu
}b_{\nu \alpha \beta },  \label{12.a}
\end{equation}

\begin{equation}
b_{\nu \alpha \beta }=u_{\nu }\delta _{\alpha \beta }-u_{\alpha }\delta
_{\nu \beta }-\frac{1}{2}u_{\lambda }e_{\lambda \beta \sigma }e_{\nu \alpha
\mu }\partial _{\mu }u_{\sigma }.  \label{12.b}
\end{equation}

Here $e_{\alpha \beta \gamma }$, $\delta _{\alpha \beta }$ are three
dimensional Levi-Civita and Kronecker symbols respectively.

Finally, with the help of (11), (12) we get the following
expressions for the nonzero components of Christoffel symbols for Euler
transformation ( \ref{8}) %
\begin{equation}
\Gamma _{00}^{\alpha }=-\frac{\ddot{u}_{\nu }}{c^{2}}S_{\alpha \nu }^{-1},%
\text{ }\Gamma _{0\beta }^{\alpha }=-\frac{(\partial _{\beta }\dot{u}_{\nu })%
}{c}S_{\alpha \nu }^{-1},\text{\ }\Gamma _{\lambda \sigma }^{\alpha
}=-(\partial _{\lambda }\partial _{\sigma }u_{\nu })S_{\alpha \nu }^{-1}%
\text{\ .}  \label{13}
\end{equation}

\section{CONTINUITY EQUATION}

For simplicity let us start from continuity equation, which in an inertial
reference frame $K^{\prime }$ has the form

\begin{equation}
\partial _{i}^{\prime }j^{\prime i}=0.  \label{14.a}
\end{equation}

After transformation ( \ref{2})  to noninertial reference frame $K$
we have covariant continuity equation: 
\begin{equation}
\frac{1}{\sqrt{-g}}\partial _{i}\left( \sqrt{-g}j^{i}\right) =0.  \label{14.b}
\end{equation}

Here $\sqrt{-g}$ is the determinant of the transformation matrix $\left(
4\right) $. On the other hand, because of the four-vector character of the
current, we have 
\begin{equation}
j^{i}\left( X\right) =\tilde{\Lambda}_{n}^{i}\left( X\right) j^{\prime
n}\left( W\left( X\right) \right) .  \label{15}
\end{equation}

By comparing (\ref{14.a})  with (\ref{14.b}) and taking
into account (\ref{15}), we may state the following. If $%
j_{0}^{i}\left( X\right) $ is a solution of the continuity equation

\begin{equation}
\partial _{i}j_{0}^{i}\left( X\right) =0,  \label{16}
\end{equation}

then

\begin{equation}
j_{1}^{i}\left( X\right) \equiv \sqrt{-g\left( X\right) }\tilde{\Lambda}%
_{n}^{i}\left( X\right) j_{0}^{n}\left( W\left( X\right) \right)  \label{17}
\end{equation}

satisfies as well the same equation (\ref{16}) for any functions $%
W^{i}\left( X\right) $.

Thus, the covariance principle provides an opportunity to construct
infinitely many solutions (\ref{17}) of continuity equation (\ref{16}) from a given one.

\section{\protect\bigskip PARTICLE CHARGE AND CURRENT DENSITIES IN THE
EXTERNAL\ FIELD}

In this Section we show that expression (\ref{17}) has very
important consequences in the electrodynamics. It provides an opportunity to
express charge and current densities in the arbitrary external field via
charge and current densities of the undisturbed system.

Let us consider an ensemble of identical particles in a medium, having
charge $e$ and trajectories $\mathbf{r}_{a}^{0}\left( t\right) $ $\left(
a=1,2,...\right) $ in the absence of external field. Then the current
four-vector at the point $X^{\prime i}\equiv \left( ct^{\prime },\mathbf{r}%
^{\prime }\right) $ is given by$^{4}$%
\begin{equation}
j_{0}^{i}\left( X^{\prime }\right) =ce\sum\limits_{a}\delta \left( \mathbf{r%
}^{\prime }-\mathbf{r}_{a}^{0}\left( t^{\prime }\right) \right) \frac{%
dX^{\prime i}}{dX^{\prime 0}}.  \label{18.a}
\end{equation}%
\ 

In the presence of an external field the trajectories of particles are
changed $\mathbf{r}_{a}^{0}\left( t\right) \rightarrow \mathbf{r}_{a}\left(
t\right) $ and the current at the point $X$\ is 
\begin{equation}
j^{i}\left( X\right) =ce\sum\limits_{a}\delta \left( \mathbf{r}-\mathbf{r}%
_{a}\left( t\right) \right) \frac{dX^{i}}{dX^{0}}.  \label{18.b}
\end{equation}%
\ 

\bigskip Now, if we perform the Euler transformation in (\ref{18.a})
and use the well-known formula%
\begin{equation*}
\prod\limits_{i=1}^{n}\delta \left( x_{i}-\alpha _{i}\right) =\frac{1}{%
\left\vert J\right\vert }\prod\limits_{i=1}^{n}\delta \left( \xi _{i}-\beta
_{i}\right) \text{,\ }J\equiv \frac{\partial \left( x_{1}....x_{n}\right) }{%
\partial \left( \xi _{1}.....\xi _{n}\right) }
\end{equation*}%
we arrive at the expression%
\begin{equation}
j^{i}\left( X\right) \equiv \sqrt{-g\left( X\right) }\tilde{\Lambda}%
_{n}^{i}\left( X\right) j_{0}^{n}\left( X-U\right) ,  \label{19}
\end{equation}

under the condition%
\begin{equation}
\mathbf{u}\left( \mathbf{r}_{a}\left( t\right) \mathbf{,}t\right) =\mathbf{r}%
_{a}\left( t\right) \mathbf{-r}_{a}^{0}\left( t\right) .  \label{20}
\end{equation}

This means that in an arbitrary external field the current four-vector $%
j^{i}\left( X\right) $ is expressed linearly in terms of the undisturbed
four-current $j_{0}^{i}$ at the point $X-U$ .

As we see, $\mathbf{u}(\mathbf{r,}t)$ is a field which on the trajectories
of particles equals to the displacement of trajectories of particles caused
by external forces and, hence, it has the analogous meaning as in the theory
of elasticity.

The expression (\ref{19}) is a special case of (\ref{17})
when $W^{i}\left( X\right) $ is the Euler transformation with the additional
condition (\ref{20}). Hence, we can say that, by using the
covariance principle, we are able to connect current and charge densities of
disturbed and undisturbed system of identical point particles in any
external field with the help of the Euler transformation.

In three-dimensional form Eq.(\ref{19}) is presented as$^{2}$%
\begin{equation}
\rho \left( \mathbf{r},t\right) =\sqrt{-g}\rho _{0}\left( \mathbf{r-u}%
,t\right) ,  \label{21.a}
\end{equation}

\begin{equation}
j_{\alpha }\left( \mathbf{r},t\right) =\sqrt{-g}S_{\alpha \beta }^{-1}[\dot{u%
}_{\beta }\rho _{0}\left( \mathbf{r-u},t\right) +j_{0\beta }\left( \mathbf{%
r-u},t\right) ],  \label{21.b}
\end{equation}

where we used (\ref{9.b}).Here $\sqrt{-g\text{ }}$\bigskip and $%
S_{\alpha \beta }^{-1}$ are given by (13), (16).

In the applications of theory we have often situations where the charges are
initially uniformly distributed and there is no current, i.e., 
\begin{eqnarray}
\rho _{0} &=&const,  \label{22} \\
\mathbf{j}_{0} &=&0.  \notag
\end{eqnarray}

In this case, with the help of (13), (16), we
can present (28), (29) in the familiar way 
\begin{eqnarray}
\rho \left( \mathbf{r,}t\right) &=&\rho _{0}-\mathbf{\nabla P}\left( \mathbf{%
r,}t\right) ,  \label{23} \\
\mathbf{j}\left( \mathbf{r,}t\right) &\mathbf{=}&\frac{\partial }{\partial t}%
\mathbf{P\left( \mathbf{r,}t\right) +}\text{ }[\mathbf{\nabla \times M}%
\left( \mathbf{r,}t\right) ].  \notag
\end{eqnarray}

Here the electric and magnetic polarization vectors have the forms 
\begin{equation}
P_{\alpha }=\rho _{0}\{u_{\alpha }+\frac{1}{2}\left[ u_{\nu }\partial _{\nu
}u_{\alpha }-u_{\alpha }\partial _{\nu }u_{\nu }\right] +\frac{1}{3}\sigma
_{\alpha \nu }u_{\nu }\},  \label{24.a}
\end{equation}

\begin{equation}
M_{\alpha }=\rho _{0}\{\frac{1}{2}e_{\alpha \lambda \nu }u_{\lambda }\dot{u}%
_{\nu }+\frac{1}{3}e_{\nu \sigma \lambda }\dot{u}_{\nu }u_{\sigma }\partial
_{\alpha }u_{\lambda }\}.  \label{24.b}
\end{equation}

Hence, in this case we can introduce electric and magnetic polarizations in
an unambiguous, natural way.

\section{BOLTZMAN\ EQUATION}

Suppose that in an inertial reference system $K^{\prime }$ the distribution
function $f^{\prime }\left( X^{\prime },P^{\prime }\right) $ of the
particles with charge $e$ and mass $m$ satisfies the relativistic Boltzman
equation$^{5}$%
\begin{equation}
P^{\prime i}\partial _{i}^{\prime }f^{\prime }\left( X^{\prime },P^{\prime
}\right) =C^{\prime }\left( X^{\prime },P^{\prime }\right)  \label{25.a}
\end{equation}

where $C^{\prime }\left( X^{\prime },P^{\prime }\right) $ is the collision
integral and $P^{^{\prime }i}$ is the momentum four-vector of particles. In
noninertial reference system $K$ we have covariant Boltzman equation$^{6}$ 
\begin{equation}
\left[ P^{i}\partial _{i}+mF^{i}\left( X,P\right) \frac{\partial }{\partial
P^{i}}\right] f\left( X,P\right) =C\left( X,P\right) ,  \label{25.b}
\end{equation}

where $F^{i}\left( X,P\right) $ is the\ inertial force given by%
\begin{equation}
F^{i}\left( X,P\right) =-\frac{1}{m}\Gamma _{jl}^{i}P^{j}P^{l}.  \label{26}
\end{equation}

Distribution function and collision integrals are scalars i.e.%
\begin{equation}
f^{\prime }\left( X^{\prime },P^{\prime }\right) =f\left( X,P\right) , 
\label{27}
\end{equation}

\begin{equation}
C^{\prime }\left( X^{\prime },P^{\prime }\right) =C\left( X,P\right) , 
\label{28}
\end{equation}

where%
\begin{equation}
X^{\prime i}=W^{i}\left( X\right) ,  \label{29.a}
\end{equation}

\begin{equation}
P^{i}=\tilde{\Lambda}_{n}^{i}P^{\prime n}.  \label{29.b}
\end{equation}

Then, from (\ref{25.a}) -(\ref{29.b}), one can say that if $%
f_{0}\left( X,P\right) $ is a solution of the free Boltzman equation

\begin{equation}
P^{i}\partial _{i}f_{0}\left( X,P\right) =C_{0}\left( X,P\right) ,  \label{30}
\end{equation}

then 
\begin{equation}
\bar{f}\left( X,P\right) =f_{0}\left( W\left( X\right) ,\check{\Lambda}%
\left( X\right) P\right)  \label{31}
\end{equation}

satisfies the Boltzman equation in the external field $F^{i}\left(
X,P\right) $ (\ref{26}) with collision integral

\begin{equation}
\bar{C}\left( X,P\right) =C_{0}\left( W\left( X\right) ,\check{\Lambda}%
\left( X\right) P\right) .  \label{32}
\end{equation}

In the last two expressions (\ref{31}) ,(\ref{32}) we have
introduced short-hand writing $\check{\Lambda}\left( X\right) P$ instead of $%
\Lambda _{m}^{i}\left( X\right) P^{m}.$

Taking into account difference of the values of phase volumes in $K^{\prime
} $ and $K$ we obtain that the distribution function corresponding to the
solution (\ref{31}) is%
\begin{equation}
f\left( X,P\right) =\left( \sqrt{-g\left( X\right) }\right) ^{2}f_{0}\left(
W\left( X\right) ,\check{\Lambda}\left( X\right) P\right) .  \label{33}
\end{equation}

As the four-current of particles is defined as 
\begin{equation}
j^{i}\left( X\right) =\frac{e}{m}\int dPP^{i}f\left( X,P\right)  \label{34}
\end{equation}

with distribution function (\ref{33}), we get, by changing
integration variables $P^{i}\rightarrow \Lambda _{m}^{i}P^{m}$, the same
expression as (\ref{17}) with%
\begin{equation}
j_{0}^{i}\left( X\right) \equiv \frac{e}{m}\int dPP^{i}f_{0}\left( X,P\right)
\end{equation}

as it must be.

Now let us consider these results for the Euler transformation in
nonrelativistic limit, i.e., for%
\begin{equation}
P^{i}=\left( mc,\mathbf{p}\right) ,  \label{35}
\end{equation}

where $\mathbf{p\equiv }m\mathbf{v}$ and $\mathbf{v}$ is the particle
velocity satisfying Newton's equation of motion:%
\begin{equation}
m\mathbf{\dot{v}}\left( t\right) =\mathbf{F.}  \label{36}
\end{equation}

Insertion of (\ref{35}) into (\ref{26}) and use of
expressions for Christoffel symbols (\ref{13}) for Euler
transformation gives the following expression for equation (\ref{25.b})

\begin{equation}
\{\frac{\partial }{\partial t}+\left( \mathbf{v\nabla }\right) +mS_{\alpha
\nu }^{-1}\left[ \ddot{u}_{\nu }+2\left( \mathbf{v\nabla }\right) \dot{u}%
_{\nu }+\left( \mathbf{v\nabla }\right) ^{2}u_{\nu }\right] \frac{\partial }{%
\partial p_{\alpha }}\}\bar{f}\left( \mathbf{r},\mathbf{p,}t\right) =C\left( 
\mathbf{r},\mathbf{p,}t\right) ,  \label{37}
\end{equation}

where

\begin{equation}
\bar{f}\left( \mathbf{r},\mathbf{p,}t\right) \equiv f_{0}\left( X-U,\check{%
\Lambda}P\right)  \label{38}
\end{equation}

\begin{equation}
C\left( \mathbf{r},\mathbf{p,}t\right) =\frac{1}{m}C_{0}\left( X-U,\check{%
\Lambda}P\right)  \label{39}
\end{equation}

On the other hand,%
\begin{equation}
\ddot{u}_{\nu }+2\left( \mathbf{v\nabla }\right) \dot{u}_{\nu }+\left( 
\mathbf{v\nabla }\right) ^{2}u_{\nu }+\left( \mathbf{\dot{v}\nabla }\right)
u_{\nu }\equiv \frac{d^{2}}{dt^{2}}u_{\nu }\left( \mathbf{r}\left( t\right)
,t\right)  \label{40}
\end{equation}

so, taking into account (\ref{36}), we have from (\ref{37})
\begin{equation}
\{\frac{\partial }{\partial t}+\left( \mathbf{v\nabla }\right) +\mathbf{F}%
\frac{\partial }{\partial \mathbf{p}}\}\bar{f}\left( \mathbf{r},\mathbf{p,}%
t\right) =C\left( \mathbf{r},\mathbf{p,}t\right)  \label{41}
\end{equation}

provided the condition

\begin{equation}
m\frac{d^{2}}{dt^{2}}\mathbf{u}\left( \mathbf{r}\left( t\right) ,t\right) =%
\mathbf{F}  \label{42}
\end{equation}

is met.

This condition is equivalent to (\ref{20}) and may be written as%
\begin{equation}
\mathbf{u}\left( \mathbf{r}\left( t\right) ,t\right) =\mathbf{r}\left(
t\right) -\mathbf{r}^{0}\left( t\right) ,  \label{43}
\end{equation}

where $\mathbf{r}^{0}\left( t\right) $ is undisturbed trajectory of the
particle ( i.e. switching of the external field $\mathbf{F}$ replaces $%
\mathbf{r}^{0}\left( t\right) $ by $\mathbf{r}\left( t\right) $).

The equation (\ref{41}) is seen to be the Boltzman equation in the
external field of the force $\mathbf{F}$ and so we can confirm once more
that having any solution of free Boltzman equation (\ref{30}) we
are able to construct a new solution (\ref{38}) for the same
equation in an arbitrary external field (\ref{41}) in
nonrelativistic limit.

Let us now suppose that the external field is switched on at $t=0.$ In this
case the problem above corresponds to the Cauchy problem for Boltzman
equation and it may be stated that we have solved the Cauchy problem for
Boltzman equation in an external field in nonrelativistic limit.

Finally, we would like to recall that the real distribution function
corresponding to the solution (\ref{38}) is given by

\begin{equation}
f\left( \mathbf{r},\mathbf{p,}t\right) =\left( \sqrt{-g}\right)
^{2}f_{0}\left( \mathbf{r-u,}\check{\Lambda}P\mathbf{,}t\right)  \label{44}
\end{equation}

as mentioned above.

\section{OPEN ELECTRODYNAMICS}

Let us start from phenomenological electrodynamics in an inertial frame $%
K^{\prime }$ with Cartesian coordinates, i.e., from Maxwell equations in
Minkowski representation 
\begin{equation}
\partial _{j}^{\prime }H^{\prime ij}=-\frac{4\pi }{c}j_{ext}^{\prime i}, 
\label{45.a}
\end{equation}

\begin{equation}
\partial _{j}^{\prime }F_{il}^{\prime }+\partial _{i}^{\prime
}F_{lj}^{\prime }+\partial _{l}^{\prime }F_{ji}^{\prime }=0,  \label{45.b}
\end{equation}

\bigskip and the most general phenomenological expansion%
\begin{equation}
H^{\prime ij}\left( X^{\prime }\right) =\sum\limits_{s=1}^{\infty }\int
dX_{1}^{\prime }..dX_{s}^{\prime }\varepsilon
^{ijl_{1}m_{1}...l_{s}m_{s}}\left( X^{\prime },X_{1}^{\prime
},..X_{s}^{\prime }\right) F_{l_{1}m_{1}}^{\prime }\left( X_{1}^{\prime
}\right) ...F_{l_{s}m_{s}}^{\prime }\left( X_{s}^{\prime }\right) . 
\label{45.c}
\end{equation}

Here $H^{ij}=\left( -\mathbf{D,H}\right) $, $F_{ij}=\left( \mathbf{E,B}%
\right) $ are Minkowski tensors of electromagnetic field, $j_{ext}^{i}$ is
the four-current of external charges, $\varepsilon
^{ijl_{1}m_{1}...l_{s}m_{s}}\left( X,X_{1},..X_{s}\right) $ is a $2(s+1)$%
-rank tensor describing electromagnetic properties of the medium (tensor of
electromagnetic permittivity).

In the frame $K,$ after transformation ( \ref{2}) , Maxwell equations 
(\ref{45.a}), (\ref{45.b}) are 
\begin{equation}
\frac{1}{\sqrt{-g}}\partial _{j}(\sqrt{-g}H^{ij})=-\frac{4\pi }{c}%
j_{ext}^{i},  \label{46.a}
\end{equation}

\begin{equation}
\partial _{j}F_{il}+\partial _{i}F_{lj}+\partial _{l}F_{ji}=0,  \label{46.b}
\end{equation}

\begin{eqnarray}
H^{ij}\left( X\right) &=&\sum\limits_{s=1}^{\infty }\int
dX_{1}..dX_{s}\varepsilon ^{ijl_{1}m_{1}...l_{s}m_{s}}\left(
X,X_{1},..X_{s}\right) \times  \label{46.c} \\
&&\times \sqrt{-g\left( X_{1}\right) }F_{l_{1}m_{1}}\left( X_{1}\right) ...%
\sqrt{-g\left( X_{s}\right) }F_{l_{s}m_{s}}\left( X_{s}\right) ,
\end{eqnarray}
where
\begin{eqnarray}
\varepsilon ^{ijl_{1}m_{1}...l_{s}m_{s}}\left( X,X_{1},..X_{s}\right) &=&
\tilde{\Lambda}_{i^{\prime }}^{i}\left( X\right) \tilde{\Lambda}_{j^{\prime
}}^{j}\left( X\right) \tilde{\Lambda}_{l_{1}^{\prime }}^{l_{1}}\left(
X_{1}\right) \tilde{\Lambda}_{m_{1}^{\prime }}^{m_{1}}\left( X_{1}\right) ...
\tilde{\Lambda}_{l_{s}^{\prime }}^{l_{s}}(X_{s})\tilde{\Lambda}
_{m_{s}^{\prime }}^{m_{s}}(X_{s})\times  \label{47} \nonumber\\
&&\times \varepsilon ^{i^{\prime }j^{\prime }l_{1}^{\prime }m_{1}^{\prime
}...l_{s}^{\prime }m_{s}^{\prime }}(W\left( X),W(X_{1}),..W(X_{s})\right) .
\end{eqnarray}
By defining%
\begin{equation}
\bar{H}^{ij}\left( X\right) =\sqrt{-g\left( X\right) }\tilde{\Lambda}%
_{m}^{i}\left( X\right) \tilde{\Lambda}_{n}^{j}\left( X\right) H^{mn}\left(
W\left( X\right) \right) ,  \label{48.a}
\end{equation}

\begin{equation}
\bar{F}_{ij}\left( X\right) =\Lambda _{i}^{m}\left( X\right) \Lambda
_{j}^{n}\left( X\right) F_{mn}\left( W\left( X\right) \right) ,  \label{48.b}
\end{equation}

\begin{equation}
\bar{j}_{ext}^{i}\left( X\right) =\sqrt{-g\left( X\right) }\tilde{\Lambda}%
_{m}^{i}\left( X\right) j_{ext}^{m}\left( W\left( X\right) \right) , 
\label{48.c}
\end{equation}

we can state, after comparing (\ref{45.a}), (\ref{45.b}) with (\ref{46.a}), (\ref{46.b}):
if $\check{H}$, $\check{F}$ tensors satisfy Maxwell equations%
\begin{equation}
\partial _{j}H^{ij}=-\frac{4\pi }{c}j_{ext}^{i}\text{, \ \ }\partial
_{j}F_{il}+\partial _{i}F_{lj}+\partial _{l}F_{ji}=0,  \label{49.a}
\end{equation}

with material equation%
\begin{equation}
H^{ij}\left( X\right) =\sum\limits_{s=1}^{\infty }\int
dX_{1}..dX_{s}\varepsilon ^{ijl_{1}m_{1}...l_{s}m_{s}}\left(
X,X_{1},..X_{s}\right) F_{l_{1}m_{1}}\left( X_{1}\right)
...F_{l_{s}m_{s}}\left( X_{s}\right) ,  \label{49.b}
\end{equation}

then (65)-(67) satisfy the same set of equations with material
equation%
\begin{equation}
\bar{H}^{ij}\left( X\right) =\sum\limits_{s=1}^{\infty }\int dX_{1}..dX_{s}%
\bar{\varepsilon}^{ijl_{1}m_{1}...l_{s}m_{s}}(X,X_{1},..X_{s})\bar{F}%
_{l_{1}m_{1}}\left( X_{1}\right) ...\bar{F}_{l_{s}m_{s}}\left( X_{s}\right) .
\label{49.c}
\end{equation}

and the electromagnetic permittivity tensor

\begin{eqnarray}
\bar{\varepsilon}^{ijl_{1}m_{1}....l_{s}m_{s}}\left( X,X_{1}....X_{s}\right)
&\equiv &\gamma _{i^{\prime }j^{\prime }}^{ij}\left( X\right) \gamma
_{l_{1}^{\prime }m_{1}^{\prime }}^{l_{1}m_{1}}\left( X_{1}\right) ...\gamma
_{l_{s}^{\prime }m_{s}^{\prime }}^{l_{s}m_{s}}\left( X_{s}\right) \times 
\label{49.d} \nonumber\\
&&\times \varepsilon ^{i^{\prime }j^{\prime }l_{1}^{\prime }m_{1}^{\prime
}...l_{s}^{\prime }m_{s}^{\prime }}\left( W(X),W(X_{1}),..W(X_{s}\right) ).
\end{eqnarray}

Here%
\begin{equation}
\gamma _{mn}^{ij}\left( X\right) \equiv \sqrt{-g\left( X\right) }\tilde{%
\Lambda}_{m}^{i}\left( X\right) \tilde{\Lambda}_{n}^{j}\left( X\right) . 
\label{50}
\end{equation}

So, having a solution of Maxwell equations in a medium with a certain
permittivity tensor $\varepsilon ^{ijl_{1}m_{1}...l_{s}m_{s}}\left(
X,X_{1},..X_{s}\right) $ we are able to construct infinitely many other
exact solutions of the Maxwell equations in the media with permittivity
tensors $\bar{\varepsilon}^{ijl_{1}m_{1}....l_{s}m_{s}}\left(
X,X_{1}....X_{s}\right) $.

In the case where the initial phenomenological equation (\ref{49.b})
is local one, i.e., if

\begin{equation}
H^{ij}\left( X\right) =\sum\limits_{s=1}^{\infty }\varepsilon
^{ijl_{1}m_{1}...l_{s}m_{s}}\left( X\right) F_{l_{1}m_{1}}\left( X\right)
...F_{l_{s}m_{s}}\left( X\right) ,  \label{51}
\end{equation}

then it is very easy to show that (\ref{48.a})-(\ref{48.c}) satisfy the Maxwell
equations with material equation%
\begin{equation}
\bar{H}^{ij}\left( X\right) =\sum\limits_{s=1}^{\infty }\bar{\varepsilon}%
^{ijl_{1}m_{1}...l_{s}m_{s}}\left( X\right) \bar{F}_{l_{1}m_{1}}\left(
X\right) ...\bar{F}_{l_{s}m_{s}}\left( X\right) ,  \label{52.a}
\end{equation}

and the electromagnetic permittivity tensor:

\begin{eqnarray}
\bar{\varepsilon}^{ijl_{1}m_{1}...l_{s}m_{s}}\left( X\right) &=&\sqrt{%
-g\left( X\right) }\tilde{\Lambda}_{i^{\prime }}^{i}\left( X\right) \tilde{%
\Lambda}_{j^{\prime }}^{j}\left( X\right) \tilde{\Lambda}_{l_{1}^{\prime
}}^{l_{1}}\left( X\right) \tilde{\Lambda}_{m_{1}^{\prime }}^{m_{1}}\left(
X\right) ...\tilde{\Lambda}_{l_{s}^{\prime }}^{l_{s}}(X)\tilde{\Lambda}%
_{m_{s}^{\prime }}^{m_{s}}(X)\times  \label{52.b}\nonumber\\
&&\times \varepsilon ^{i^{\prime }j^{\prime }l_{1}^{\prime }m_{1}^{\prime
}...l_{s}^{\prime }m_{s}^{\prime }}(W\left( X)\right) .
\end{eqnarray}

\section{LOCAL LINEAR PROBLEMS}

In this case (\ref{51}) has the following appearance:

\begin{equation}
H^{ij}\left( X\right) =\varepsilon ^{ijmn}\left( X\right) F_{mn}\left(
X\right) ,  \label{53}
\end{equation}

and for (\ref{52.b}) we have%
\begin{equation}
\bar{\varepsilon}^{ijlm}\left( X\right) =\sqrt{-g\left( X\right) }\tilde{%
\Lambda}_{i^{\prime }}^{i}\left( X\right) \tilde{\Lambda}_{j^{\prime
}}^{j}\left( X\right) \tilde{\Lambda}_{l^{\prime }}^{l}\left( X\right) 
\tilde{\Lambda}_{m^{\prime }}^{m}\left( X\right) \varepsilon ^{i^{\prime
}j^{\prime }l^{\prime }m^{\prime }}(W\left( X)\right) .  \label{54}
\end{equation}

In three-dimensional notations (\ref{53}) becomes%
\begin{eqnarray}
D_{\alpha } &=&-2\varepsilon ^{0\alpha 0\beta }E_{\beta }+\varepsilon
^{0\alpha \beta \gamma }e_{\beta \gamma \sigma }B_{\sigma },  \label{55} \\
H_{\alpha } &=&\frac{1}{2}e_{\alpha \beta \gamma }\varepsilon ^{\beta \gamma
\lambda \mu }e_{\lambda \mu \sigma }B_{\sigma }-e_{\alpha \beta \gamma
}\varepsilon ^{\beta \gamma 0\sigma }E_{\sigma }.  \notag
\end{eqnarray}

For ordinary media%
\begin{eqnarray}
D_{\alpha } &=&\varepsilon _{\alpha \beta }E_{\beta },  \label{56} \\
B_{\alpha } &=&\mu _{\alpha \beta }H_{\beta }.  \notag
\end{eqnarray}

Then (\ref{55}), (\ref{56}) give for nonzero components of 
$\varepsilon ^{\beta \gamma \lambda \mu }$ 
\begin{eqnarray}
\varepsilon ^{0\alpha \beta 0} &=&\frac{1}{2}\varepsilon _{\alpha \beta }, 
\label{57} \\
\varepsilon ^{\alpha \beta \lambda \mu } &=&\frac{1}{2}e_{\alpha \beta \nu
}\mu _{\nu \sigma }^{-1}e_{\sigma \lambda \mu }.  \notag
\end{eqnarray}

Let's consider time independent transformation $\left( 2\right) $, i.e.,%
\begin{equation}
\mathbf{r}^{\prime }=\mathbf{W}\left( \mathbf{r}\right) \mathbf{,}\text{ }%
t^{\prime }=t;  \label{58}
\end{equation}

then for nonzero components of transformation matrices $\Lambda $, $\tilde{%
\Lambda}$ we have from $\left( 4\right) $%
\begin{equation}
\Lambda _{0}^{0}=\tilde{\Lambda}_{0}^{0}=1,\text{\ }\Lambda _{\beta
}^{\alpha }=\partial _{\beta }W_{\alpha }\equiv M_{\alpha \beta },\text{\ }%
\tilde{\Lambda}_{\beta }^{\alpha }=M_{\alpha \beta }^{-1},\text{\ }\sqrt{-g}%
=\left\Vert M\right\Vert  \label{59}
\end{equation}

and, hence, for the new dielectric and magnetic tensors we obtain from 
(\ref{54}), (\ref{57}) and (\ref{59}) %
\begin{equation}
\bar{\varepsilon}_{\alpha \beta }\left( \mathbf{r}\right) =\left\Vert
M\right\Vert M_{\alpha \nu }^{-1}M_{\beta \lambda }^{-1}\varepsilon _{\nu
\lambda }\left( \mathbf{W}\left( \mathbf{r}\right) \right) ,  \label{60.a}
\end{equation}

\begin{equation}
\bar{\mu}_{\alpha \beta }\left( \mathbf{r}\right) =\left\Vert M\right\Vert
M_{\alpha \nu }^{-1}M_{\beta \lambda }^{-1}\mu _{\nu \lambda }\left( \mathbf{%
W}\left( \mathbf{r}\right) \right) .  \label{60.b}
\end{equation}

From (\ref{48.a})-(\ref{48.c}) , (\ref{59}) we find the solutions of Maxwell
equations in a medium with $\bar{\varepsilon}_{\alpha \beta }\left( \mathbf{r%
}\right) ,\bar{\mu}_{\alpha \beta }\left( \mathbf{r}\right) $ as:

\begin{equation}
\bar{E}_{\alpha }\left( \mathbf{r,}t\right) =M_{\beta \alpha }E_{\beta
}\left( \mathbf{W}\left( \mathbf{r}\right) ,t\right)  \label{61.a}
\end{equation}

\begin{equation}
\bar{B}_{\alpha }\left( \mathbf{r,}t\right) =\left\Vert M\right\Vert
M_{\alpha \beta }^{-1}B_{\beta }\left( \mathbf{W}\left( \mathbf{r}\right)
,t\right)  \label{61.b}
\end{equation}

in the presence of the following external sources:%
\begin{equation}
\bar{\rho}_{ext}\left( \mathbf{r,}t\right) =\left\Vert M\right\Vert \rho
_{ext}\left( \mathbf{W}\left( \mathbf{r}\right) ,t\right) ,  \label{62.a}
\end{equation}

\begin{equation}
\lbrack \bar{j}_{ext}\left( \mathbf{r,}t\right) ]_{\alpha }=\left\Vert
M\right\Vert M_{\alpha \beta }^{-1}[j_{ext}\left( \mathbf{W}\left( \mathbf{r}%
\right) ,t\right) ]_{\beta }.  \label{62.b}
\end{equation}

As an illustration of our results let us construct a new solution of Maxwell
equations in a medium if we have these solutions in vacuum $\left(
j_{ext}\equiv 0\right) $.

For the vacuum%
\begin{equation}
\varepsilon _{\alpha \beta }=\mu _{\alpha \beta }=\delta _{\alpha \beta } 
\label{63}
\end{equation}

and, with the help of transformation (\ref{58}) and 
(\ref{60.a}), (\ref{60.b}) ,(\ref{63}) , we come to a medium with equal dielectric
and magnetic permittivity tensors:%
\begin{equation}
\bar{\varepsilon}_{\alpha \beta }\left( \mathbf{r}\right) =\bar{\mu}_{\alpha
\beta }\left( \mathbf{r}\right) =\left\Vert M\right\Vert M_{\alpha \nu
}^{-1}M_{\beta \nu }^{-1}.  \label{64}
\end{equation}

As the simplest case of transformation (\ref{58}) let us take%
\begin{equation}
x^{\prime }=x,\text{ \ }y^{\prime }=y,\text{ \ }z^{\prime }=f\left( z\right)
,\text{ \ }t^{\prime }=t  \label{65}
\end{equation}

where $f\left( z\right) $ is an arbitrary monotonically increasing function%
\begin{equation}
f^{\prime }\left( z\right) \equiv n\left( z\right) >0.  \label{66}
\end{equation}

From (\ref{59}) we have for nonzero matrix elements of $\hat{M}$ 
\begin{equation*}
M_{xx}=M_{yy}=1,\text{ \ }M_{zz}=n\left( z\right) ,\text{ \ }
\end{equation*}

and (\ref{64}) gives%
\begin{equation}
\bar{\varepsilon}_{xx}=\bar{\varepsilon}_{yy}=\bar{\mu}_{xx}=\bar{\mu}%
_{yy}=n\left( z\right) ,\text{ \ \ }\bar{\varepsilon}_{zz}=\bar{\mu}_{zz}=%
\frac{1}{n\left( z\right) }  \label{67}
\end{equation}

From (\ref{61.a}),(\ref{61.b}) we have for the solutions of Maxwell equations in
the medium with equal electrical and magnetic tonsorial permittivity
(\ref{67})%
\begin{eqnarray}
\bar{E}_{\alpha }\left( \mathbf{r,}t\right) &=&E_{\alpha }\left( x,y,f\left(
z\right) ,t\right) ,\text{ \ }\alpha =x,y  \label{68.a} \\
\bar{E}_{z}\left( \mathbf{r,}t\right) &=&n\left( z\right) E_{z}\left(
x,y,f\left( z\right) ,t\right)  \notag
\end{eqnarray}

\begin{eqnarray}
\bar{B}_{\alpha }\left( \mathbf{r,}t\right) &=&n\left( z\right) B_{\alpha
}\left( x,y,f\left( z\right) ,t\right) ,\text{ \ }\alpha =x,y  \label{68.b}
\\
\bar{B}_{z}\left( \mathbf{r,}t\right) &=&B_{z}\left( x,y,f\left( z\right)
,t\right) ,  \notag
\end{eqnarray}

where $\mathbf{E}\left( \mathbf{r},t\right) ,$ $\mathbf{B}\left( \mathbf{r}%
,t\right) $ are any solutions of Maxwell equation in vacuum (for instance,
harmonic functions).

Up to now we knew solution of Maxwell equations for media with refractive
index with constant, linear, quadratic and exponential coordinate
dependence. Our approach extends these solutions to anisotropic media with
primitivities containing arbitrary function (\ref{67}).

\section{ELECTRON PLASMA ON THE BACKGROUND OF IONS IN A SLAB}

In this section we intend to show how to use the developed technique in
construction of a solution for the self-consistent problem of plasma
interacting with electromagnetic field (EMF).

Consider two component plasma (electrons and ions). For simplicity suppose
that ions are very heavy and interaction with EMF does not change their
state of equilibrium. Besides, let us also assume that by some reason they
cannot leave the borders of a slab $\left[ -a\leq x\leq a\right] $. Then, we
come to the problem of electron plasma on the positively charged background
in a slab.

The charge density in the slab of positive background is:%
\begin{equation}
\rho _{0}\left( x\right) =-en\left[ \theta \left( x+a\right) -\theta \left(
x-a\right) \right] ,  \label{69}
\end{equation}

where $n$ is the concentration of ions. Then, in the presence of external
EMF for electron plasma we have Maxwell-Boltzman equations:%
\begin{equation}
\partial _{j}F^{ij}=-\frac{4\pi }{c}(j^{i}+j_{0}^{i}),  \label{70.a}
\end{equation}

\begin{equation}
\partial _{j}F_{il}+\partial _{i}F_{lj}+\partial _{l}F_{ji}=0,  \label{70.b}
\end{equation}

\begin{equation}
\left[ P^{i}\partial _{i}+\frac{e}{mc}F^{ij}\left( X\right) P_{j}\frac{%
\partial }{\partial P^{i}}\right] f\left( X,P\right) =C\left( X,P\right) . 
\label{70.c}
\end{equation}

Here the electron current four-vector is defined as%
\begin{equation}
j^{i}\left( X\right) =\frac{e}{m}\int dPP^{i}f\left( X,P\right) ,  \label{71.a}
\end{equation}

while the ion current four-vector is

\begin{equation}
j_{0}^{i}\left( x\right) \equiv \left( c\rho _{0}\left( x\right)
,0,0,0\right) .  \label{71.b}
\end{equation}

Let us search for the solution of (\ref{70.a})-(\ref{70.c}) as:%
\begin{equation}
F^{ij}=F_{ion}^{ij}+F_{e}^{ij},  \label{72}
\end{equation}

where $F_{ion}^{ij}$ is the static field of ions having the charge
distribution (\ref{69}) in the form 
\begin{equation}
F_{ion}^{10}=-F_{ion10}=-4\pi en\left[ x\theta \left( a-\left\vert
x\right\vert \right) +a\theta \left( \left\vert x\right\vert -a\right) %
\right] .  \label{73}
\end{equation}

Substituting (\ref{72}) into (\ref{70.a})-(\ref{70.c}) results in 
\begin{equation}
\partial _{j}F_{e}^{ij}=-\frac{4\pi }{c}j^{i}  \label{74.a}
\end{equation}

\begin{equation}
\partial _{j}F_{eil}+\partial _{i}F_{elj}+\partial _{l}F_{eji}=0,  \label{74.b}
\end{equation}

\begin{equation}
\left[ P^{i}\partial _{i}+\frac{e}{mc}(F_{e}^{ij}+F_{ion}^{ij})P_{j}\frac{%
\partial }{\partial P^{i}}\right] f\left( X,P\right) =C\left( X,P\right) . 
\label{74.c}
\end{equation}

Now our problem is to construct the solution for (\ref{74.a})-(\ref{74.c}).

Let us start from the equations:%
\begin{equation}
\partial _{j}F^{\left( 0\right) ij}=-\frac{4\pi }{c}j^{\left( 0\right) i}, 
\label{75.a}
\end{equation}

\begin{equation}
P^{i}\partial _{i}f^{\left( 0\right) }=C_{0}.  \label{75.b}
\end{equation}

Here $f^{\left( 0\right) }$ is the electron distribution function in
equilibrium which satisfies the neutrality condition 
\begin{equation}
j^{\left( 0\right) i}=\frac{e}{m}\int dPP^{i}f^{\left( 0\right) }\equiv
-j_{0}^{i}\left( x\right) .  \label{76}
\end{equation}

Hence, as a solution of (\ref{75.a}) we can take 
\begin{equation}
F^{\left( 0\right) ij}=-F_{ion}^{ij}+\bar{F}^{ij},  \label{77}
\end{equation}

where $\bar{F}^{ij}$ is a solution of Maxwell equations in vacuum.

After the Euler transformation in (\ref{75.a}), (\ref{75.b}) with $\mathbf{u}\left( 
\mathbf{r,}t\right) $ satisfying the equation%
\begin{equation}
\frac{d^{2}}{dt^{2}}u_{\nu }\left( \mathbf{r}\left( t\right) ,t\right) =%
\frac{e}{m^{2}c}F^{\nu j}\left( \mathbf{r}\left( t\right) ,t\right) P_{j}, 
\label{78}
\end{equation}

the equation (\ref{75.a}) becomes%
\begin{equation}
\partial _{j}(\sqrt{-g}F^{ij})=-\frac{4\pi }{c}j^{i},  \label{79.a}
\end{equation}

where%
\begin{equation}
F^{ij}\left( X\right) =\tilde{\Lambda}_{m}^{i}\tilde{\Lambda}%
_{n}^{j}F^{\left( 0\right) mn}\left( X-U\right) ,  \label{79.b}
\end{equation}

\begin{equation}
j^{i}\left( X\right) =\sqrt{-g}\tilde{\Lambda}_{m}^{i}j^{\left( 0\right)
m}\left( X-U\right) .  \label{79.c}
\end{equation}

Besides, after this transformation (\ref{75.b}) goes to
(\ref{74.c}) with electron distribution function 
\begin{equation}
f\left( X,P\right) =\left( \sqrt{-g}\right) ^{2}f^{\left( 0\right) }\left(
X-U,\check{\Lambda}P\right) .  \label{80}
\end{equation}

Because of (\ref{79.b}) solution of (\ref{79.a}) is%
\begin{equation}
\tilde{F}^{ij}\left( X\right) =\sqrt{-g}\tilde{\Lambda}_{m}^{i}\tilde{\Lambda%
}_{n}^{j}F^{\left( 0\right) mn}\left( X-U\right) .  \label{81}
\end{equation}

Then, with the help of the\ formula (\ref{A.7}) of the Appendix and 
(\ref{72}) ,(\ref{81}) , we come to the following solution for
EMF%
\begin{equation}
F^{ij}\left( X\right) =F_{ion}^{ij}\left( X\right) +\tilde{F}^{ij}\left(
X\right) +\frac{1}{2}e^{ijpq}e_{slmn}\partial _{p}\int dX_{1}G_{q}^{s}\left(
X,X_{1}\right) \partial _{1}^{l}\tilde{F}^{mn}\left( X_{1}\right) .  \label{82}
\end{equation}

So, the solutions of Maxwell-Boltzman equations, (\ref{70.a})-(\ref{70.c}) , for
electron plasma on the positive background in a slab in the external EMF is
given by Eq.(\ref{80}) for electron distribution function and by Eq.%
(\ref{82}) for EMF tensor.

\begin{center}
\textbf{SUMMARY}
\end{center}

With the help of the Einstein covariance principle we succeeded in:

\textbf{a.} constructing infinitely many solutions to the continuity
equation, Boltzman equation and Maxwell phenomenological equations, if we
have some single solution of these equations.

\textbf{b.} obtaining a general expression for the charge and current
densities of the system of charged point particles in an arbitrary external
field.

\textbf{c.} solving the Cauchy problem for the nonrelativistic Boltzman
equation in an arbitrary external field.

\textbf{d.} solving the Maxwell equations in an anisotropic inhomogenious
medium with equal electrical and magnetic primitivities.

\textbf{e.} demonstrating an algoritm for construction of a solution of the
self-consistent problem of the interaction of electron plasma with the
external electromagnetic field on the slab of positive ions.

\begin{center}
\textbf{ACKNOWLEDGMENTS}
\end{center}

Authors would like to thank Prof. V.O.Chaltykyan, for critically reading of
manuscript and for valuable remarks.

This work was supported by the SCOPES Swiss grant 7UKPJ062150.

\begin{center}
\textbf{APPENDIX}
\end{center}

Let $\tilde{F}^{ij}$ be an antisymmetric tensor satisfying the first Maxwell
equation:%
\begin{equation}
\partial _{j}\tilde{F}^{ij}=-\frac{4\pi }{c}j^{i}\text{.}  \label{A.1}
\end{equation}

\textbf{Problem} is in construction of an antisymmetric tensor $F^{ij}$
satisfying both Maxwell equations having $\tilde{F}^{ij}$.

\textbf{Solution: }Because $\tilde{F}^{ij}$ satisfy (\ref{A.1}) ,
for an arbitrary four-vector $A^{i}$\ 
\begin{equation}
F^{ij}\left( X\right) =\tilde{F}^{ij}\left( X\right) +e^{ijpq}\partial
_{p}A_{q}\left( X\right)   \label{A.2}
\end{equation}

is as well a solution of (\ref{A.1}) .

Let us claim (\ref{A.2})  to be a solution of the second Maxwell
equation, i.e.,%
\begin{equation}
e_{ijkl}\partial ^{j}F^{kl}\left( X\right) =0.  \label{A.3}
\end{equation}

From (\ref{A.2}) , (\ref{A.3}) , we get equation for $A_{j}$:%
\begin{equation}
\left[ \partial _{i}\partial ^{j}-\delta _{i}^{j}\square _{X}\right]
A_{j}\left( X\right) =\partial ^{j}\tilde{F}_{ij}^{\ast }\left( X\right) , 
\label{A.4.a}
\end{equation}

where $\tilde{F}_{ij}^{\ast }$ is dual tensor of $\tilde{F}_{ij}$%
\begin{equation}
\tilde{F}_{ij}^{\ast }=\frac{1}{2}e_{ijmn}\tilde{F}^{mn}.  \label{A.4.b}
\end{equation}

Let $G_{j}^{l}\left( X,X^{\prime }\right) $ be the Green's function of the
wave equation%
\begin{equation}
\left[ \partial _{i}\partial ^{j}-\delta _{i}^{j}\square _{X}\right]
G_{j}^{l}\left( X,X^{\prime }\right) =\delta _{i}^{l}\delta \left(
X-X^{\prime }\right) .  \label{A.5}
\end{equation}

Then, taking $A_{j}$ as a solution of inhomogenious equation 
(\ref{A.4.a}) ,%
\begin{equation}
A_{j}\left( X\right) =\int dX_{1}G_{j}^{i}\left( X,X_{1}\right) \partial
_{1}^{l}\tilde{F}_{il}^{\ast }\left( X_{1}\right)   \label{A.6}
\end{equation}

we will satisfy the second Maxwell equation for the tensor $F^{ij}$ given by 
(\ref{A.2}) .

Hence, from (\ref{A.2}) ,(121), (122)  and (\ref{A.6}) 
we finally get the expression for $F^{ij}$ satisfying both Maxwell equations

\begin{center}
\begin{equation}
F^{ij}\left( X\right) =\tilde{F}^{ij}\left( X\right) +\frac{1}{2}%
e^{ijpq}e_{slmn}\partial _{p}\int dX_{1}G_{q}^{s}\left( X,X_{1}\right)
\partial _{1}^{l}\tilde{F}^{mn}\left( X_{1}\right) .  \label{A.7}
\end{equation}

\textbf{REFERENCES}
\end{center}

$^{1}$A. Einstein, Ann.Phys.\textbf{\ 49}, 769 (1916).

$^{2}$V.M. Mekhitarian, (unpublished) (1982).

$^{3}$V.E. Mkrtchian, R.v. Baltz, J. Math. Phys. \textbf{41}, 1956 (2000).

$^{4}$L.D. Landau, E.M. Lifshitz, "The classical theory of fields" (Nauka,
Moscow. 1988).

$^{5}$S.R. de Groot, W.A. van Leeuwen, Ch.G. van Weert, "Relativistic
kinetic theory" (North-Holland Publishing \ \ \ \ \ Company, Amsterdam.
1980).

$^{6}$A.A. Vlasov, "Statistical distribution functions" (Nauka, Moscow.
1966).

\end{document}